\documentclass{emulateapj}
\accepted{to ApJ}

\shortauthors{Matthews \& Dupree}
\shorttitle{Spatially Resolved VLA Observations of Betelgeuse}

\begin{document}
\newcommand{\ang}{\rm \AA}
\newcommand{\msun}{M$_\odot$}
\newcommand{\lsun}{L$_\odot$}
\newcommand{\days}{$d$}
\newcommand{\degree}{$^\circ$}
\newcommand{\ud}{{\rm d}}
\newcommand{\as}[2]{$#1''\,\hspace{-1.7mm}.\hspace{.0mm}#2$}
\newcommand{\am}[2]{$#1'\,\hspace{-1.7mm}.\hspace{.0mm}#2$}
\newcommand{\ad}[2]{$#1^{\circ}\,\hspace{-1.7mm}.\hspace{.0mm}#2$}
\newcommand{\lsim}{~\rlap{$<$}{\lower 1.0ex\hbox{$\sim$}}}
\newcommand{\gsim}{~\rlap{$>$}{\lower 1.0ex\hbox{$\sim$}}}
\newcommand{\HA}{H$\alpha$}
\newcommand{\HII}{\mbox{H\,{\sc ii}}}
\newcommand{\kms}{\mbox{km s$^{-1}$}}
\newcommand{\HI}{\mbox{H\,{\sc i}}}
\newcommand{\CII}{\mbox{C\,{\sc ii}}}
\newcommand{\MgII}{\mbox{Mg\,{\sc ii}}}
\newcommand{\jks}{Jy~km~s$^{-1}$}

\title{Spatially Resolved Observations of Betelgeuse at $\lambda$7~mm
  and $\lambda$1.3~cm Just Prior to the Great Dimming}

\author{L. D. Matthews\altaffilmark{1} \&  A. K. Dupree\altaffilmark{2}}

\altaffiltext{1}{Massachusetts Institute of Technology 
Haystack Observatory, 99 Millstone Road, Westford, MA
  01886 USA}
\altaffiltext{2}{Center for Astrophysics $|$ Harvard \& Smithsonian, 60
  Garden Street, MS-15, Cambridge, MA 02138 USA}

\begin{abstract}
We present spatially resolved observations of Betelgeuse ($\alpha$~Orionis) obtained 
with the Karl G. Jansky Very Large Array (VLA) at 
$\lambda\sim$7~mm (44~GHz) and $\lambda\sim$1.3~cm (22~GHz) on 2019
August 2, just prior to the onset of the historical optical dimming
that occurred between late 2019 and early 2020. Our measurements suggest recent changes in the temperature and density
structure of the atmosphere between radii $r\sim 2$--$3R{\star}$. 
At $\lambda$7~mm the star is $\sim$20\% dimmer than in previously published observing epochs between 1996--2004.
We measure a mean gas temperature
of $T_{B}=2270\pm260$~K at $r\sim2.1R{\star}$, where $R_{\star}$ is the canonical
photospheric radius. This is $\sim2\sigma$ lower than previously reported temperatures at comparable radii
and $>$1200~K lower than predicted by previous semi-empirical
models of the atmosphere. The measured brightness temperature at $r\sim2.6R_{\star}$
($T_{B}=2580\pm 260$~K) is also cooler than expected
based on trends in past measurements. The stellar brightness profile in our current measurements appears relatively smooth
  and symmetric,
  with no obvious signatures of giant convective cells or other surface features. However,
  the azimuthally averaged brightness profile is found to be more complex than a uniform elliptical
  disk. Our
observations were obtained approximately six weeks before 
spectroscopic measurements in the ultraviolet revealed evidence of
increases in the chromospheric electron density in the southern
hemisphere of Betelgeuse, coupled with a
large-scale outflow. We discuss possible scenarios linking these events with the observed
radio properties of the star, including the passage of a strong shock wave.

\end{abstract}

\keywords{Stellar chromospheres (230); Stellar photospheres (1237);
   Red Giant stars (1372); Stellar properties (1624); Stellar surfaces
  (1632)}  

\section{Introduction\protect\label{Intro}}
The red supergiant Betelgeuse ($\alpha$~Orionis) has been the recent
subject of intense observational scrutiny owing to
an unprecedented optical dimming of the star observed in late 2019 and early
2020 (see Calderwood 2021). 
While the visible light curve of Betelgeuse is known to
undergo semi-regular variations in optical brightness
on timescales of $\sim$300-500 days and $\sim$2000 days, respectively
(e.g., Kiss et al. 2006), the ``Great Dimming'' of 2019/2020 (which reached its brightness minimum
from 2020 February 7--13; Guinan et al. 2020) marked the faintest appearance of
the star in nearly 200 years of recorded photometry. 
Two hypotheses that emerged to
explain the dramatic dimming are recent dust formation or a reduction in
photospheric temperature---or perhaps a combination of these effects. 

Based on optical spectrophotometric measurements of TiO from
2020 February, Levesque \& Massey (2020)
argued that the
dimming of the star
could not be explained by a temperature decrease alone. They found $T_{\rm eff}=3600\pm25$~K, only
  marginally cooler than seen in measurements made in 2004 ($T_{\rm eff}=3650\pm$25~K).
These authors proposed instead episodic mass loss, coupled with an
increase in large-grain circumstellar dust along the
line-of-sight. However, using higher resolution spectra of atomic lines, Za\v{c}s \&
Puk\={i}tis (2021) found evidence for increased macroturbulence during
the Great Dimming, coupled with a statistically significant
temperature decrease to $T_{\rm eff}\approx3500$~K.
From the analysis of Wing TiO and near-infrared photometry spanning five years,
Harper et al. (2020)
argued that no new dust is required, and that the dimming can be explained by the presence of photospheric
inhomogeneities with a large covering fraction ($\ge$50\%) and a mean effective
temperature$\ge$250~K cooler than expected for an M2 Iab supergiant. A similar scenario
was suggested
by Dharmawardena et al. (2020), who  measured  a flux decrease of $\sim$20\% at submillimeter
wavelengths during the Great Dimming
compared with measurements between 2007--2017 and concluded that the observed flux change
could not
be explained by dust, but may result either from 
changes in the temperature and/or radius of
the star. Based on aperture polarimetry, Cotton et al. (2020) found that polarization appeared during
the Great Dimming that could be explained by  photospheric asymmetries and/or obscuration by grains.
Meanwhile George et al. (2020) have argued that the Great Dimming may have been linked with a
critical shift in the pulsation dynamics of Betelgeuse.

Several key insights into the behavior of Betelgeuse near
the time of the Great Dimming were
provided by spatially resolved ultraviolet (UV) spectroscopy from the {\it
  Hubble Space Telescope} ({\it HST}), spanning several epochs
in 2019 and 2020 (Dupree et al. 2020). Measurements of the \MgII\ $h$ and $k$ lines revealed evidence for the passage
of a shock or pressure  wave through the southwestern portion of the star's atmosphere
between 2019 September and November, a time
when the photosphere was at its maximum outflow velocity (relative to
the mean) as a result of the phase of the 400-day pulsation cycle (see
Figure~2b of Dupree et al.).
Analysis of the \CII\ line revealed that the passage of this
wave was accompanied by increases in the temperature and electron
density in the southern hemisphere.
A scenario proposed by Dupree et al. is that an
exceptionally strong convective
upwelling from the photosphere (enhanced in strength by the phase of the pulsation
cycle) led to a major ejection event that launched material through the
chromosphere, beyond which it may have cooled sufficiently to allow
the formation of dust.
High spatial resolution optical images obtained by Montarg\`es et al. (2021) appear to
reinforce this picture, showing that the southern hemisphere of the star was ten times
darker during the Great Dimming compared with images obtained a year earlier. The authors
attribute the observed darkening to obscuration by a newly formed dust clump. Using a
tomographic analysis, Kravchenko et al. (2021) found evidence for successive shock waves
that could explain the outflow detected by Dupree et al. (2020), but argued that an
increase in molecular opacity rather than dust formation best explains the subsequent dimming.

Past 
studies of red supergiants at radio (centimeter, millimeter, and submillimeter)
wavelengths have shown that they are sources of
continuum emission that is predominantly free-free in origin. At wavelengths longer than a few millimeters, a
significant
fraction of their continuum emission appears to arise from a component of chromospheric
gas at cooler temperatures than the material that emits in the
UV, and with a larger filling factor (Lim et al. 1998; Harper, Brown, \& Lim 2001; O'Gorman et
al. 2020). At submillimeter wavelengths, blackbody emission from the stellar disk becomes increasingly
dominant.  Because the radio emission is
thermal and optically thick, the Rayleigh-Jeans limit applies to the
radiative transfer equation, and
radio observations with sufficient angular
resolution to spatially resolve
the stars permit direct measurements of the stellar diameter and the mean
gas (electron) temperature (e.g., Reid \& Menten 1997; Lim et al. 1998; Harper et
al. 2001; O'Gorman et al. 2017;
see also Section~\ref{temperature}). For these reasons, radio observations
provide a valuable complement to UV measurements for the
study of red supergiant chromospheres and the adjacent layers of the atmosphere.

In conjunction with our ongoing multi-cycle {\it HST} program (Dupree 2018)
we were  awarded a
single epoch of observations of Betelgeuse with the Karl G. Jansky Very Large Array
(VLA)\footnote{The VLA of the National Radio Astronomy
  Observatory (NRAO) is operated by Associated
  Universities, Inc. under cooperative agreement with the National
  Science Foundation.} at wavelengths of $\lambda$1.3~cm and $\lambda$7~mm.  These two wavelengths are of
particular interest with respect to the UV observations of Dupree et al. (2020), since they
probe gas at radii comparable to those giving rise the UV continuum, and just interior to that region,
respectively (Lim et al. 1998). Here we present the results of these new VLA
observations, which were carried out on 2019
August 2, just prior to the Great Dimming and the appearance of the outflow event reported by
Dupree et al. (2020).

\section{Previous Radio Observations of Betelgeuse\protect\label{previous}}
Betelgeuse is classified as an M2Iab supergiant of variability class
SRc (Kiss et al. 2006). Its proximity
(222$^{+40}_{-34}$~pc; Harper et al. 2017) makes the angular size of its
photosphere 
the largest of any star visible from the northern hemisphere
($\theta_{\star}$=44.2$\pm$0.2 mas at $\lambda$2.2$\mu$m; Dyck et al. 2002). 

Betelgeuse has been the target of radio wavelength studies for more
than half a century (Kellermann \& Pauliny-Toth 1966; Seaquist
1967). Early hints that the radio emission has a significant
chromospheric contribution came from the multi-frequency study of
Altenhoff, Oster, \& Wendker (1979).
The first observations of Betelgeuse with the VLA were published by Newell
\& Hjellming (1982) at $\nu$=1.46, 4.89, 15.0, and 22.5~GHz. Although these latter
measurements did not spatially resolve the star,
the inferred spectral index $\alpha$=1.32 (where flux density $S_{\nu} \propto \nu^{\alpha}$)
 led the
authors to conclude that
the radio emission must be thermal in origin, and they suggested that it arises predominantly from an optically
thick chromosphere with an extent of a few stellar radii.

The first spatially resolved radio observations of Betelgeuse were
obtained by
Skinner et al. (1997)  using the
Multi-Element Radio Linked Interferometer Network (MERLIN) and the VLA
at $\lambda$6~cm, confirming that the radio emission is extended to
several times the photospheric diameter, consistent with a
chromospheric origin. 
Subsequently
Lim et al. (1998) were able to spatially resolve Betelgeuse at several wavelengths
between 7~mm and 6~cm using the VLA. Those observations revealed the temperature
structure of the atmosphere to be complex and established
that shorter radio wavelengths sample material at
successively smaller radii. At $r\sim2R_{\star}$ (which corresponds to
unity optical depth 
at $\lambda$7~mm) they measured a
brightness (electron) temperature of 3450$\pm$850~K, roughly consistent with
the photospheric temperature; however
the temperatures were seen to steadily decrease with larger $r$, reaching
1370$\pm$330~K at $r\sim7R_{\star}$. These temperatures are
well below the values of $\ge$4000--8000~K inferred from 
measurements of chromospheric tracers in the
optical and UV across similar radii (e.g., Gilliland \& Dupree 1996; Dupree et
al. 2020),
leading Lim et al. to conclude that the hot chromospheric material
probed by the UV must have a relatively small filling factor. The 7~mm image
presented by Lim et al. also implied that radio emission from Betelgeuse was highly
asymmetric, which the authors attributed to the
possible presence of giant convective cells.

Harper et al. (2001) used  the
observations of Lim et al. (1998)  to develop a semi-empirical
model for the extended atmosphere of Betelgeuse with a mean chromospheric temperature of 3800~K, significantly
lower than other models (cf. Hartmann \& Avrett 1984). An updated version of this model was recently
discussed by O'Gorman et al. (2020), but across the range of radii of relevance for comparsion with our
VLA measurements, the difference is negligible (G. Harper, private communication). The Harper et al. model
includes a region between
$r\sim$1.6--1.9$R_{\star}$ where the temperature dips below the mean
effective temperature.\footnote{The values quoted here have been radially scaled by a factor
  of (56/44) to account for
the difference in angular diameter of Betelgeuse's photosphere adopted in the present paper.}
 The presence of this
temperature minimum was  first indicated by infrared observations
(Tsuji 2000, 2006) and was further confirmed  by O'Gorman et
al. (2017) using spatially resolved submillimeter (338~GHz) observations from the Atacama
Large Millimeter/submillimeter Array (ALMA).
The ALMA data of O'Gorman et
al. (2017) also showed evidence of inhomogeneities in the atmosphere which the authors
suggested may arise from localized heating. 

The temporal behavior of Betelgeuse's radio emission
has been examined by 
O'Gorman et al. (2015) using resolved imaging from the VLA and the
Pie Town antenna of the Very Long Baseline Array at wavelengths
ranging from 7~mm to 6~cm.  These authors found evidence for 
 modest flux density variations with time ($\lsim\pm$25\%; see also
 Drake et al. 1992) that appear
 to mirror the star's cyclic $V$ magnitude variations---except at 7~mm where
 no evidence of variations was seen to within measurement
 uncertainties. In addition to the cyclic variations, O'Gorman et al. found 
that the flux density between $\lambda$1.3--6~cm decreased overall by
$\sim$20\%  between the 1970s and 1980s and the early 2000s, despite maintaining a
roughly constant spectral index.

%
\begin{figure}
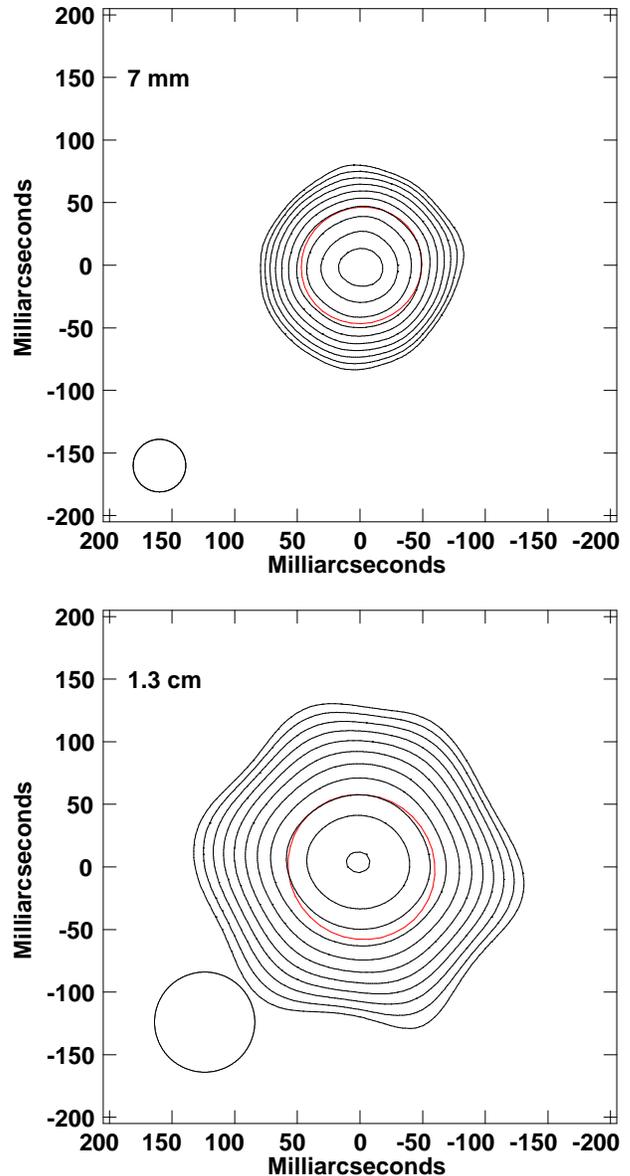

  \scalebox{0.45}{\rotatebox{0}{\includegraphics{f1a.ps}}}
  \scalebox{0.45}{\rotatebox{0}{\includegraphics{f1b.ps}}}
\caption{ {\sc CLEAN} contour maps of Betelgeuse obtained
  on 2019 August 2 at 7~mm (44~GHz; left) and 1.3~cm (22~GHz; right). Contour levels are
  ($-$10[absent], 10, 14.1, 20, 28.3, 40, 56.6, 80, 113.1, 160, 200)$\times29.1\mu$Jy
  beam$^{-1}$ (44~GHz) and
  ($-$10[absent], 10, 14.1, 20, 28.3, 40, 56.6, 80, 113.1, 160, 226, 319)$\times15.7\mu$Jy
  beam$^{-1}$ (22~GHz) where the lowest contours are
  10$\sigma$. The adopted restoring beam is shown in the lower left of
  each panel (see Table~2). The red ellipses indicate the best-fitting uniform elliptical disk
models (Table~3).}  
\label{fig:cleanmaps}
\end{figure}

\section{Observations\protect\label{observations}}
On 2019 August 2
we carried out new continuum observations of Betelgeuse 
at 44~GHz (Q-band; $\lambda\sim7$~mm) and 22~GHz
(K-band; $\lambda\sim1.3$~cm) using the VLA in its most extended (A)
configuration.
Antenna separations ranged from 0.68~km
to 36.4~km. At $\lambda$7~mm, this configuration provided angular resolution 
of $\sim$42~mas, comparable to the size of Betelgeuse's disk in the near infrared (see Section~2).
At $\lambda$1.3~cm the angular resolution was $\sim$80~mas.

The Q-band observations were obtained using the 3-bit observing
mode and dual circular polarizations. The WIDAR correlator was
configured with four baseband pairs tuned to contiguously cover
a frequency window of $\sim$7.9~GHz, centered near 44~GHz. 
Each baseband pair contained either 15 or 16
subbands, each of which had a bandwidth of 128~MHz and 128 spectral
channels. An analogous setup was used for the K-band observations, but
with a center frequency of $\sim$22~GHz. To improve $u$-$v$ coverage, observations in the two bands were interleaved during a
  4.0 hour session, spanning approximately 14:45--18:45~UTC. Total integration time on Betelgeuse was $\sim$62~min in Q band and $\sim$42~min
  in K band and the observed elevation of the star ranged from 43~deg to 63~deg.
Data were recorded with 2-second time resolution.

The observations were carried out during the late morning and early afternoon local time.
Weather conditions ranged from clear to partly cloudy with
wind speeds $\le$4.2~m s$^{-1}$.  The RMS atmospheric phase fluctuations reported
  at the VLA site were 6--7~deg during the first half of the observations and 12--17~deg
  during the second half.
Antenna pointing corrections were evaluated hourly
using observations of a strong point source at X-band (8~GHz).  
3C48 was observed in both K and Q band to allow calibration of the absolute
flux density scale (see Section~4).
To calibrate the
atmospheric phases, fast switching was used between
Betelgeuse and two neighboring gain calibrators: J0532+0732 (``cal 1'') and
J0552+0313 (``cal 2''), which lie at projected separations of
\ad{5}{58} and \ad{4}{22}, respectively, from
the star. For Q-band, the adopted observing
sequence was: [$t_{\rm cal 1}, (t_{\rm star}, t_{\rm cal 2})\times4,
t_{\rm star}, t_{\rm cal 1}...$], where $t_{\rm cal 1}=$ 44 sec, $t_{\rm star}= $46
sec, and $t_{\rm cal
  2}= $38 sec. For K-band the sequence was: [$t_{\rm cal 1}, (t_{\rm star}, t_{\rm cal 2})\times2,
t_{\rm star}, t_{\rm cal 1}...$], where $t_{\rm cal 1}=$ 42 sec, $t_{\rm star}=$ 132
sec, and $t_{\rm cal
  2}=$ 36 sec.

\section{Data Reduction\protect\label{reduction}}
Data reduction and calibration were performed using the Astronomical Image Processing
System (AIPS; Greisen 2003). 
The original archival
science data model 
files were loaded into AIPS using the Obit software package (Cotton
2008). However, the default calibration (`CL') table was 
regenerated  to  update the gain and
  opacity information, and antenna positions were updated to the best
  available values.

After 
flagging visibly corrupted data, a requantizer gain correction was
applied using the AIPS task {\small\sc{TYAPL}}. Instrumental delays
were corrected via fringe fitting to a 1-minute segment
    of 3C48 data, with separate delay solutions
  determined for each of the four independent basebands.
Bandpass calibration was performed in the standard manner using 
J0532+0732 as the calibrator;
a spectral index of $\alpha\approx$0 was assumed (Healey et
al. 2007).

The flux density calibrator 3C48 was known to be undergoing a flare
during 2019\footnote{\url{https://science.nrao.edu/facilities/vla/docs/manuals/oss/ performance/fdscale}};
therefore special care was taken with the
absolute flux density calibration of our data. Clean component models of 3C48
appropriate for the
two observing bands were obtained from NRAO via anonymous
ftp.\footnote{ftp.aoc.nrao.edu, directory pub/staff/rperley/MODELMAPS2019.}
The models were derived by R. Perley from 
observations of 3C48 in the VLA A configuration in 2019
October and were used to set the absolute
flux density scale, assuming the standard coefficients of Perley \& Butler
(2017). We found that the differences in the derived flux density scales
based on the 2019 October Perley models  compared with the
default (non-flaring) models for 3C48 were $\sim$2\% in Q band and $\sim$3\% in K
band, respectively. Thus uncertainty in the absolute
flux calibration scale resulting from the flare is expected to be
quite small.

As a check that
the 2019 October Perley model is suitable for calibrating our
measurements, we imaged 3C48 using our new
data and found the source structure to be virtually indistinguishable
from that seen in the Perley model images, consistent with minimal change in the
source at K and Q band frequencies between the date of our
observations (2019 August 2) and the date of Perley's observations (2019
October 24). As a final check on the robustness of our flux density scale, we
compared our VLA measurements of the bandpass/complex gain calibrator
J052+0732 with a 91.5~GHz measurement from the ALMA obtained on 2019 July 26 with an independent absolute flux density
calibrator. Because of the flat spectral index of the source,
we expect similar flux densities at 91.5~GHz, 44~GHz, and 22~GHz, and
consistent with this, the reported ALMA  flux density  was
1.22$\pm$0.06~Jy\footnote{Value taken from the 
ALMA Calibrator Source Catalogue;
https://almascience.nrao.edu/sc/.}, in good agreement with our VLA
measurements (Table~1). 

%
\begin{deluxetable*}{lrrcc}
\tabletypesize{\footnotesize}
\tablewidth{0pc}
\tablenum{1}
\tablecaption{Calibration Sources}
\tablehead{
\colhead{Source} & \colhead{$\alpha$(J2000.0)} &
\colhead{$\delta$(J2000.0)} & \colhead{Flux Density (Jy)} &
\colhead{$\nu$ (GHz)} 
}
\startdata
3C48$^{a}$ & 01 37 41.2994 & 33 09 35.133 & 0.6134$^{*}$ & 44.0 \\
... & ... & ... & 1.2194$^{*}$ & 22.0 \\

J0532+0732$^{b}$ & 05 32 38.9985 & 07 32 43.346 & 1.131$\pm$0.034 & 44.0\\
      ...        &     ...       &  ...         & 1.172$\pm$0.009 & 22.0   \\ 

J0552+0313$^{c}$ & 05 52 50.1015 & 03 13 27.243 & 0.569$\pm$0.020 & 44.0 \\
    ...         & ...           & ...           & 0.734$\pm$0.006 & 22.0 \\
\enddata

\tablecomments{Units of right ascension are hours, minutes, and
seconds, and units of declination are degrees, arcminutes, and
arcseconds. Explanation of columns: (1) source name; (2) \& (3) right
ascension and declination (J2000.0); (4) derived flux density in Jy at
the frequency indicated in column 5;
(5) frequency at which the flux density in
the fourth column was computed.}

\tablenotetext{*}{Adopted value, calculated at the frequency in
  column~5 using the 2019 October clean component models from
  R. Perley and the
  coefficients from Perley \& Butler (2017; see Section~4).}
\tablenotetext{a}{Flux density calibrator.}
\tablenotetext{b}{Complex gain and bandpass calibrator.}
\tablenotetext{c}{Complex gain calibrator.}

\end{deluxetable*}

Calibration of the
frequency-independent portion of the complex gains was performed in the
standard manner using the measurements of
the gain calibrators J0532+0732 and  J0552+0313.  
Initially, phase-only corrections were solved for using these sources
and applied to Betelgeuse using linear interpolation in time. This was
followed by similarly computed
amplitude and phase corrections. The rms
scatter in the phase solutions for the two calibrators was $\lsim$5~deg in Q band and $\lsim$2~deg in K
band, respectively, for all antennas.

Following application of the computed gain solutions to Betelgeuse, an
additional iteration of phase-only self-calibration was performed for
each of the bands
using the emission from Betelgeuse itself, with 30~s solution
intervals. 
This step resulted in improvements in the rms
noise of of images of 18\% and 37\%, respectively, for Q and K bands, and the correlated
amplitude of the star increased by 52\% and 19\%, in the two bands,
respectively, following the self-calibration.

At this stage, optimized weights for the visibility data were
calculated using the AIPS task {\small\sc{REWAY}}. Finally, the
data in each band were
averaged in frequency to produce 8 spectral channels in each of the 62 subbands.

Images of Betelgeuse were produced from the fully calibrated data using
{\small\sc{CLEAN}} deconvolution  as implemented in AIPS. Robust weighting with {\cal R}=0 was
adopted, along with a circular restoring beam (see Table~2). 
Images of the star at the two observing wavelengths are shown in Figure~\ref{fig:cleanmaps}.
The images appear relatively smooth and symmetric at both wavelengths
(but see also Section~\ref{results}). Note that the apparent
six-pointed shape seen in
the outer contours of the 1.3~cm image in Figure~\ref{fig:cleanmaps}
is an artifact caused by imperfect deconvolution of the 
VLA's dirty beam, which has a six-pointed pattern whose first sidelobes
overlap with the outskirts of the stellar emission distribution in this band.


%
\begin{deluxetable*}{lcccrcc}
\tabletypesize{\footnotesize}
\tablewidth{0pc}
\tablenum{2}
\tablecaption{CLEAN Image Properties}
\tablehead{
\colhead{Band} &  \colhead{$\nu_{\rm GHz}$} &
\colhead{$\theta_{\rm a}$} & \colhead{$\theta_{\rm b}$} & \colhead{PA} &
\colhead{$\theta_{\rm circ}$} & \colhead{RMS noise}\\
\colhead{} & \colhead{(GHz)} & \colhead{(mas)} & \colhead{(mas)} &
\colhead{(degrees)}& \colhead{(mas)} & \colhead{($\mu$Jy beam$^{-1}$)}
}

\startdata
Q & 44 & 43.8 & 40.7 & $+13.7$ & 42.0 & 29.0\\
K & 22 & 87.0 & 73.8 & $+10.2$ & 80.0 & 15.9 \\
\enddata

\tablecomments{{\sc{CLEAN}} images were produced using robust  (${\cal
    R}$=0) weighting. $\nu_{\rm GHz}$ is the center frequency in GHz.
$\theta_{\rm a}$ and $\theta_{\rm b}$ are the FWHM
major and minor axes, respectively, of the dirty beam. The position
angle (PA) of the dirty beam was measured east from
  north.
The images presented in Figure~\ref{fig:cleanmaps}
were produced using circular restoring beams with the FWHM diameter $\theta_{\rm circ}$
(column~6); these values represent the geometric mean
  of the dirty beam dimensions in columns 3 \& 4.}

\end{deluxetable*}


%
\begin{deluxetable*}{lcrrrrrrr}
\tabletypesize{\footnotesize}
\tablewidth{0pc}
\tablenum{3}
\tablecaption{Measured Stellar Parameters}
\tablehead{
\colhead{Band} &  \colhead{$\nu$} & \colhead{$\theta_{\rm
    maj}$} & \colhead{$\theta_{\rm min}$} &
\colhead{PA} & \colhead{$e$} & \colhead{$S_{\nu}$} &
\colhead{$D$} & \colhead{$T_{b}$}\\
\colhead{} & \colhead{(GHz)} &  \colhead{(mas)} & \colhead{(mas)} &
\colhead{(degrees)} & \colhead{} 
& \colhead{(mJy)}    & \colhead{(AU)} & \colhead{(K)}\\
  \colhead{(1)} & \colhead{(2)} & \colhead{(3)} & \colhead{(4)} & \colhead{(5)}
& \colhead{(6)} & \colhead{(7)} & \colhead{(8)} & \colhead{(9)}  }

\startdata
\tableline
\\
\multicolumn{9}{c}{Uniform Elliptical Disk Fits to
  Visibility Data}\\
\\
\tableline
Q & 43.9 & 96.9$\pm$2.9 (0.2) & 92.6$\pm$2.8 (0.2) & $111.0\pm$5.1 (1.2) & 0.04$\pm$0.04 &
22.34$\pm$2.41 (0.03) &21.1 & 2270$\pm$260\\

K & 22.0 & 118.8$\pm$5.9 (0.3) & 114.5$\pm$5.7 (0.3) & 54.7$\pm$5.2 (1.5) & 0.04$\pm$0.07
& 9.64$\pm$0.97 (0.01)
&25.9 & 2580$\pm$260\\

\tableline \\
\multicolumn{9}{c}{Elliptical Gaussian Fits to Image Data}\\
\\
\tableline
Q & 43.9 & 66.8$\pm$2.1 (0.5) & 63.6$\pm$2.0 (0.5) & 111.0$\pm$7.8 (6.0) & 0.05$\pm$0.04 &
23.0$\pm$2.5 (0.01) & ... & ...\\

K & 22.0 & 79.4$\pm$4.0 (0.6) & 73.4$\pm$3.7 (0.6) & 67.5$\pm$6.6 (4.3) &0.07$\pm$0.07  & 9.63$\pm$1.04 (0.01)
& ... & ...\\
\enddata

\tablecomments{Quoted
  uncertainties include formal, systematic, and calibration errors
  (see Text for details) but not uncertainties in the stellar
  distance. For measured quantities the value in
  parentheses indicates the error budget contribution  from
  formal fitting uncertainties.  For a uniform elliptical disk,
$\theta_{\rm maj}$ and $\theta_{\min}$ are the major and minor axis
dimensions, respectively; for a Gaussian fit they represent the
the FWHM dimensions of the elliptical Gaussian after deconvolution from the dirty
beam.  Explanation of columns: (1) observing band; (2) mean observing frequency; (3) FWHM
  diameter of the major axis in 
mas; (4) FWHM
  diameter of the minor axis in mas; (5) PA of the major axis in
  degrees, measured east from north; (6) ellipticity, defined as 
  $e=(\theta_{\rm maj} - \theta_{\rm min})/(\theta_{\rm maj})$; (7) 
flux density in mJy; (8) mean size in AU, derived using the geometric mean angular
  diameter and assuming a distance of 222~pc; (9) brightness
  temperature in Kelvin, as derived assuming a uniform elliptical
  disk morphology (see Equation~1). }

\end{deluxetable*}

\section{Results\protect\label{results}}
\subsection{Measured Stellar Parameters\protect\label{params}}
To characterize the size and flux density of Betelgeuse in our two
VLA observing bands 
we performed fits of a two-dimensional (2D) uniform elliptical disk to
the visibility data using the AIPS {\small\sc{OMFIT}} task. As a
consistency check, we also fitted 
elliptical Gaussian models to {\small\sc{CLEAN}} deconvolved images (see
Table~2 and Figure~\ref{fig:cleanmaps})
 using the AIPS {\small\sc
  {JMFIT}} task. Results are summarized in Table~3. 
The quoted uncertainties include contributions from  the formal fitting uncertainties
(Condon 1997), as well as from calibration and
systematic errors  (see Matthews, Reid, \& Menten 2015 for details). The dominant source of uncertainty in the
derived flux densities is the  absolute
calibration uncertainty, assumed to be 10\%
in both observing
bands.\footnote{\url{https://science.nrao.edu/facilities/vla/docs/manuals/oss/ performance/fdscale.}}

At both observed wavebands the angular diameter measurements are consistent to within uncertainties with
previous spatially resolved observations of Betelgeuse
at comparable wavelengths (Lim et al. 1998; O'Gorman et al. 2015).
As expected for optically thick emission,
we measure a larger angular size for the star at the longer of our two
observing wavelengths. The mean
shape of the radio emitting surface was nearly circular during the
epoch of our observations, with an
ellipticity of at most a few per cent. The mean angular diameters 
(taken as the geometric mean of the major and minor axes measured from
the visibility data) are 94.7~mas at $\lambda$7~mm and 116.6~mas at $\lambda$1.3~cm,
respectively, corresponding to $\sim2.1$ and
$\sim2.6$ times the near-infrared diameter of the star (see Section~2).
For comparison, the angular extent of chromospheric emission traced by the UV ``continuum'' (centered near
$\lambda$2500~\ang)
is approximately 125$\pm$5~mas, or $\sim2.8R_{\star}$ in radial units (Gilliland \& Dupree 1996)\footnote{Near
  this wavelength the UV ``continuum'' from Betelgeuse
  is dominated by a
  blend of numerous
  emission lines (see e.g., Figure~10 of Brandt et al. 1995) and its angular extent will be dependent
  on the exact filter passband used.}, 
while chromospheric \MgII\ emission lines are detected out to projected radii of $\sim$48--135~mas
($\sim$1.1--6$R_{\star}$; Uitenbroek et al. 1998; Dupree et al. 2020).

While the measured angular size of Betelgeuse at both $\lambda$7~mm and $\lambda$1.3~cm is consistent with
past observations, the total flux densities in both 
observing bands are notably lower than previously published values.
Among six published $\lambda$1.3~cm measurements between 1996--2004 (Lim et al. 1998; O'Gorman et al. 2015), only 
one (8.96$\pm$0.24~mJy; obtained in 2002 April) was comparably faint
as our current measurement.  O'Gorman et al. (2015) found that the 1.3~cm
flux density appears to be correlated with the optical ($V$) magnitude, and
the 2002 measurement corresponded to a phase when the star
was near its minimal optical brightness  during its 400-day pulsation
cycle ($\phi\approx$0.5). However, at the time of
our latest measurements, Betelgeuse was at an intermediate phase  ($\phi\approx$0.09).
There is no optical photometry available during a window
from approximately 100 days preceding our VLA measurements on 2019 August 2 until 10 days after that date,
since the star was a daytime object for ground-based observers. 
Based on an extrapolation of the light curve presented by Dupree et al. (2020),
the estimated $V$ magnitude is $\sim$0.65, which is within the typical range for this phase.
To our knowledge, no published 1.3~cm measurements are
available between 2005--2019, so the more recent behavior of the star at this wavelength is
unknown.

At $\lambda$7~mm, the discrepancy between our new measurement and previous results is even more pronounced.
Based on four measurements taken between 1996 and 2004, the 7~mm flux density of Betelgeuse remained
nearly constant (28.4$\pm0.5$~mJy, where the error bar indicates the measurement dispersion;
Lim et al. 1998; O'Gorman et
al. 2015). However, the 7~mm flux density
we measure in 2019 August is $\sim$20\% 
(2$\sigma$) lower, even after accounting for calibration
uncertainties.

O'Gorman et al. (2015) noted that despite evidence for a systematic dimming of Betelgeuse at radio wavelengths,
its spectral index
has remained effectively constant over several decades: $\alpha=1.33\pm0.02$.
Based on 
our current measurements we find $\alpha=1.22\pm0.22$. Thus to within uncertainties
the spectral index remains consistent with previous measurements,
although it is
poorly constrained since we have only two observing bands.

\subsection{Brightness Temperature\protect\label{temperature}}
Because the radio emission from Betelgeuse
is optically thick and our VLA observations are spatially resolved,
it is possible to use the angular sizes
and flux density measurements from Table~3 to derive a brightness
temperature that provides a direct measurement of the mean gas
(electron) temperature via the Rayleigh-Jeans expression:
\begin{equation}
T_{B}=2S_{\nu}c^2(4.25\times10^{-28})/(k\nu^{2}\pi\theta_{\rm
    maj}\theta_{\rm min})
\end{equation}

\noindent  where $S_{\nu}$ is in mJy, $\theta_{\rm maj}$ and $\theta_{\rm min}$ are the dimensions of the
best-fitting disk model in units of mas  (see Table~3),
  $c$ is the speed of light in cm s$^{-1}$, $k$ is
  the Boltzmann constant in cgs units, and $\nu$ is the mean
  observing frequency in GHz.
At $\lambda$7~mm we find $T_{B}=2270\pm260$~K (corresponding to $r\sim2.1R_{\star}$)
and at  $\lambda$1.3~cm $T_{B}=2580\pm$260~K (corresponding to
$r\sim 2.6R_{\star}$).
The derived brightness temperature at $\lambda$1.3~cm is at the low end of previously
observed values at this wavelength (O'Gorman et
al. 2015) and is
markedly cooler than the value of $\sim$3300~K reported at
this wavelength by Lim et al. (1998). At $\lambda$7~mm, the $T_{B}$ value measured in
2019 August is the lowest ever reported at this wavelength. In
comparison, Lim et
al. (1998),  measured a brightness temperature at $\lambda$7~mm
of 3450$\pm$850~K (based on data from 1996),  comparable
to the mean photospheric temperature of Betelgeuse, while O'Gorman et
al. (2015) reported values of 3040$\pm$80~K,  2760$\pm$200~K, 2940$\pm$140~K
 for data obtained in 2000, 2003, and 2004, respectively. The semi-empirical
model of Harper et al. (2001) (which was originally based on the Lim et
al. measurements) predicts a characteristic temperature at $r\sim
2.1R_{\star}$
of $\sim$1500~K hotter than seen in our current data (see Figure~\ref{fig:toymodel}, discussed below).
Indeed, the temperature that we measure based on our new 7~mm data is only slightly warmer than
that within the molecular layer known as the ``MOLsphere", which extends from
$r\sim$1.3--1.5$R_{\star}$ and has an estimated temperature of $\sim$1500-2000~K
(Tsuji 2000, 2006).
We discuss the implications of these temperature measurements further in
Section~\ref{discussion}.

\subsection{Deviations from a Uniform Disk\protect\label{deviation}}
\subsubsection{Evidence for a Multi-component Brightness Profile\protect\label{multicomp}}
The uniform elliptical disk fits we have employed above
are useful for providing a characterization of the mean properties of the
star (angular size, temperature). However, the high signal-to-noise ratio of our current
data---roughly an order of magnitude higher than previous radio images of Betelgeuse at
comparable wavelengths---allows us to see clear evidence of deviation
 from this simple model.

\begin{figure}
  \centering
  \scalebox{0.45}{\rotatebox{0}{\includegraphics{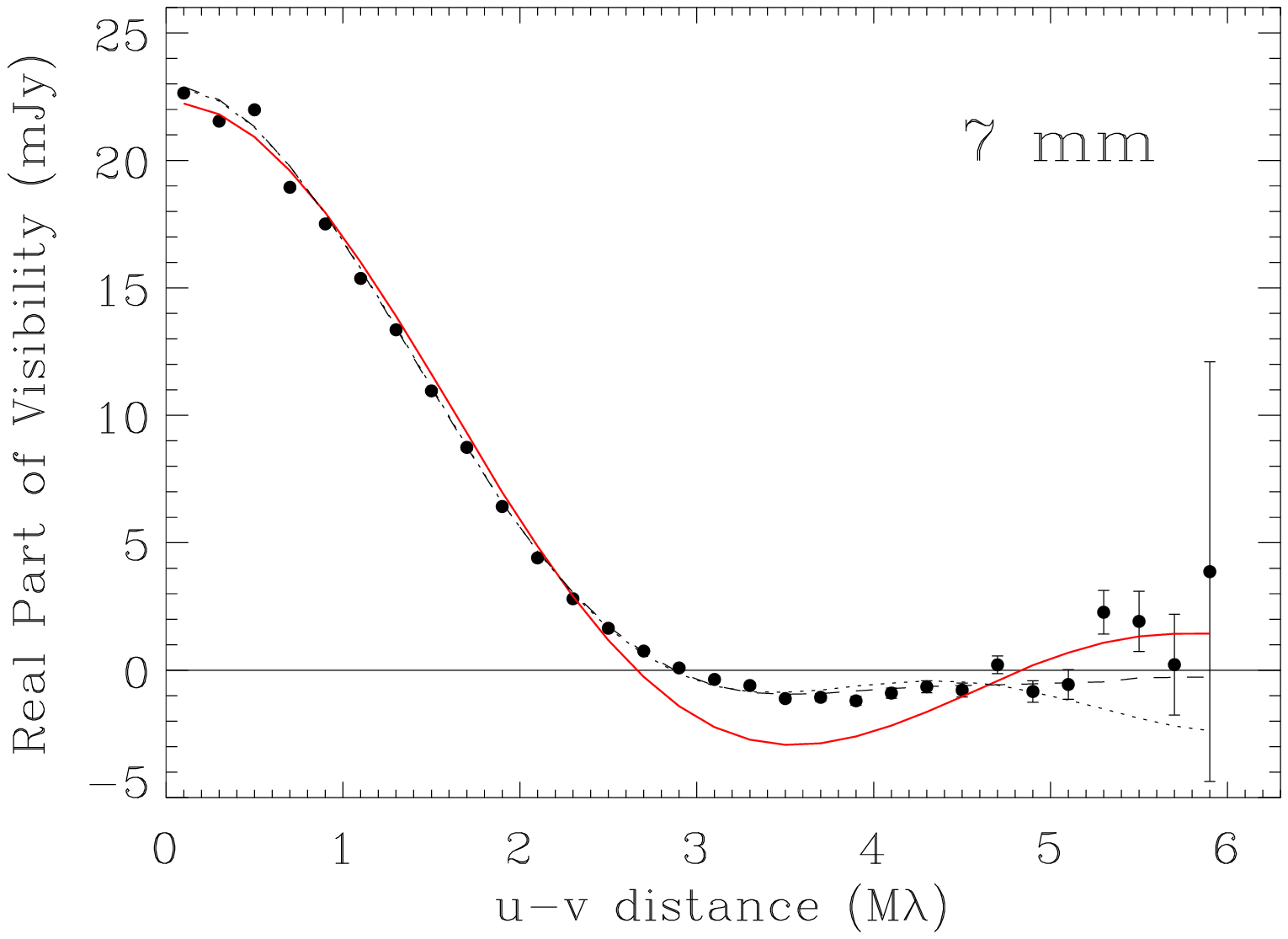}}}
  \scalebox{0.45}{\rotatebox{0}{\includegraphics{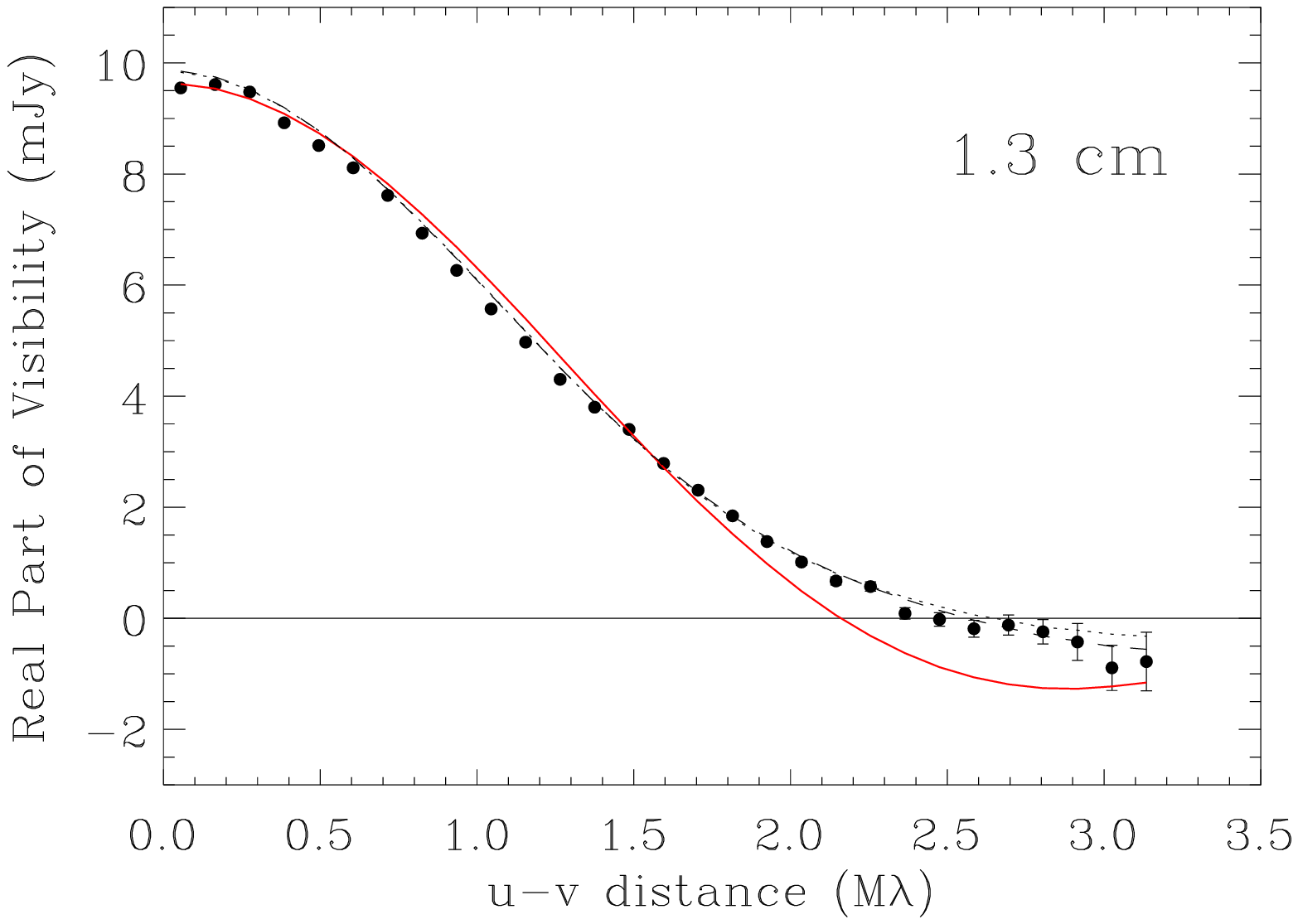}}}
  \scalebox{0.45}{\rotatebox{0}{\includegraphics{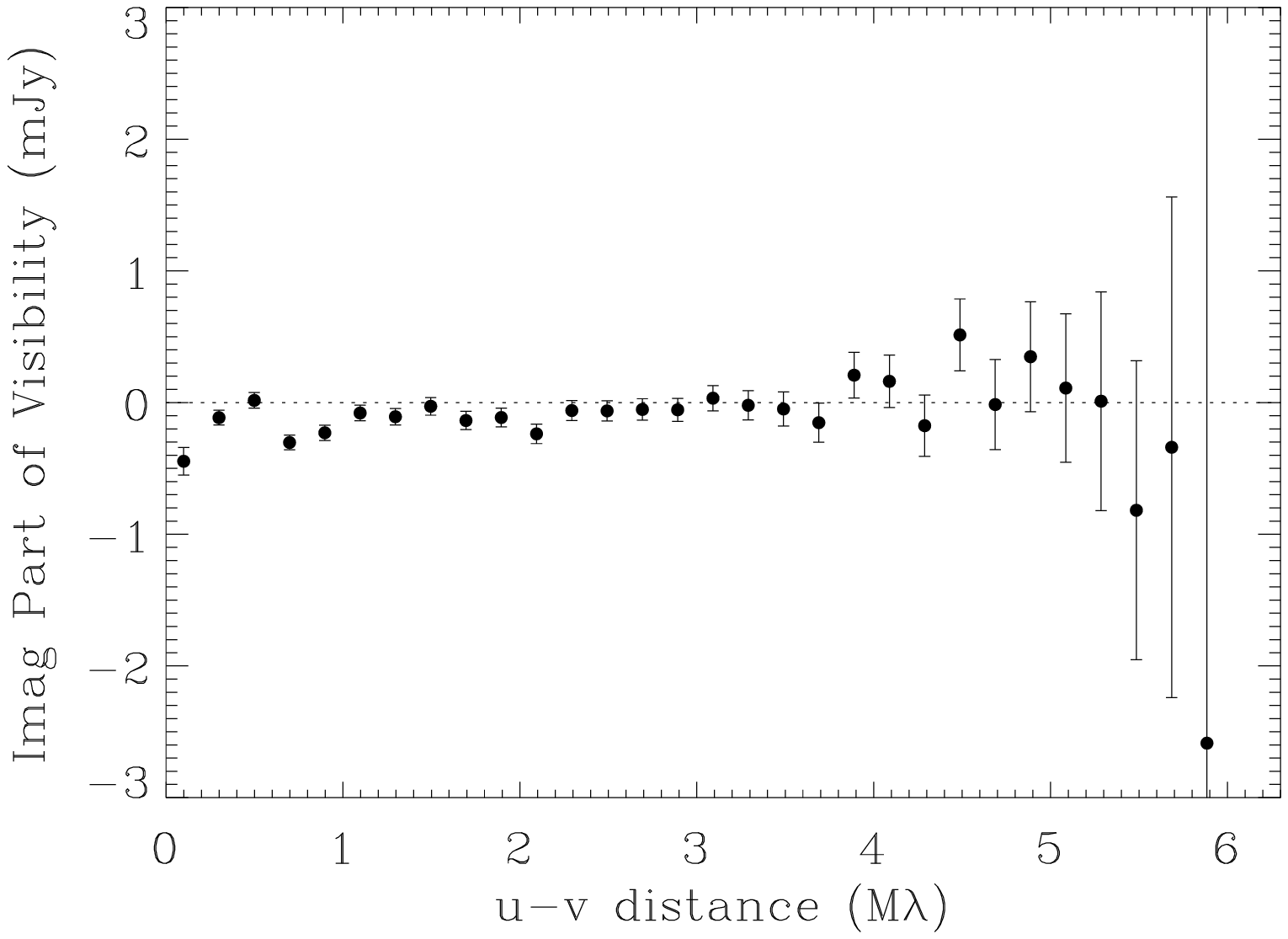}}}
  \scalebox{0.45}{\rotatebox{0}{\includegraphics{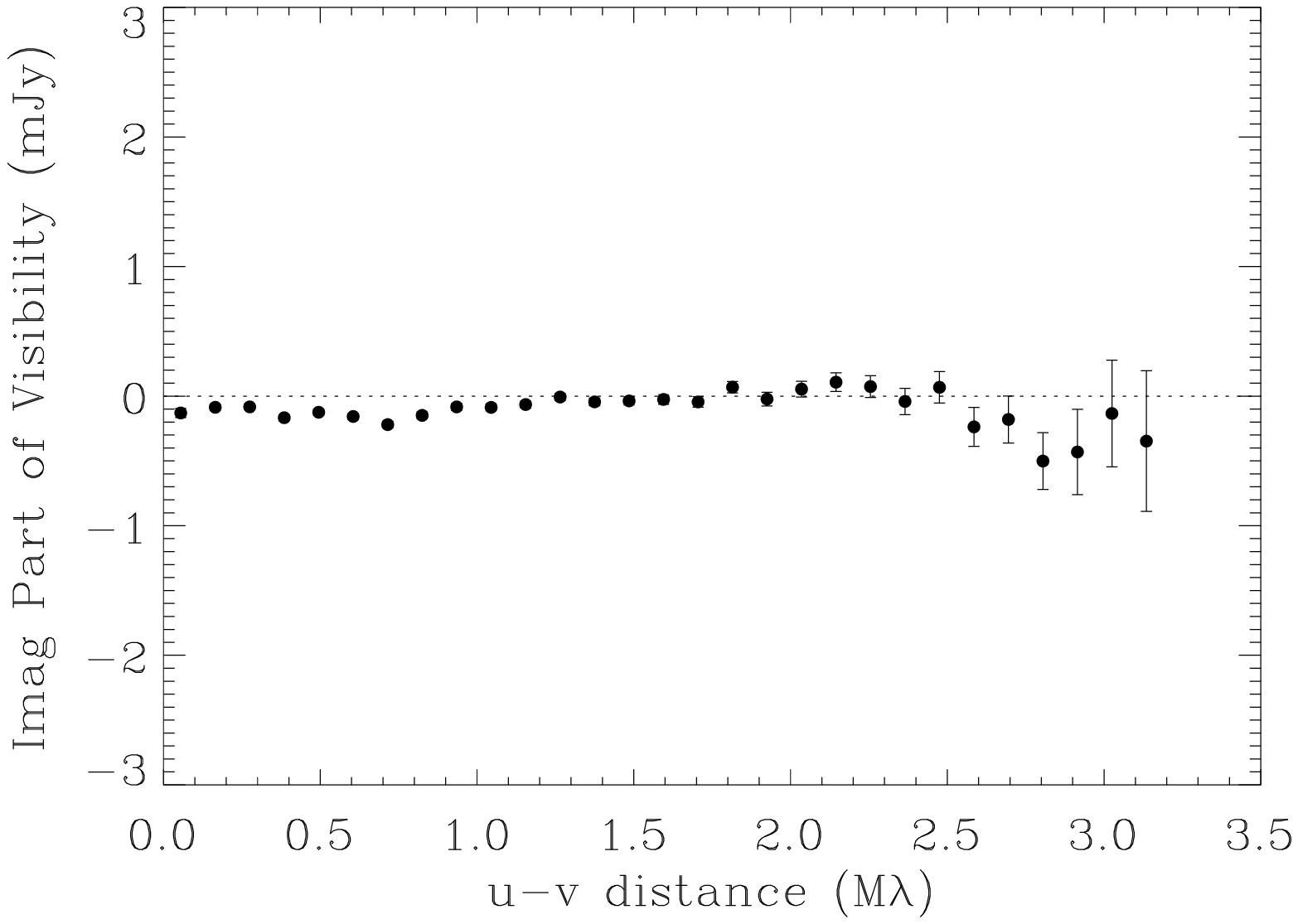}}}
  \caption{Real (top) and imaginary (bottom) parts of the visibility
    amplitude vs. baseline length for observations of Betelgeuse at
    $\lambda$7~mm (left) and $\lambda$1.3~cm (right). On the upper panels the solid red lines indicate
    the
    best-fitting uniform elliptical disk models from Table~3. The
    two-component models from Table~4 are overplotted as
    black dashed lines (Model 1) and dotted lines (Model 2). }  
\label{fig:visplots}
\end{figure}

In Figure~\ref{fig:visplots} we plot the azimuthally averaged
visibility data (real and imaginary parts) as a function of $u$-$v$ distance for our two observing
bands, with the best-fitting
uniform elliptical disk models overplotted. In each band we see that
the uniform elliptical disk model 
under-represents the observed flux on the smallest observed angular scales
(longest baselines) while systematically slightly over-predicting
the flux density on scales sampled by intermediate baselines ($\sim$0.5--1.5M$\lambda$). 

\begin{figure}
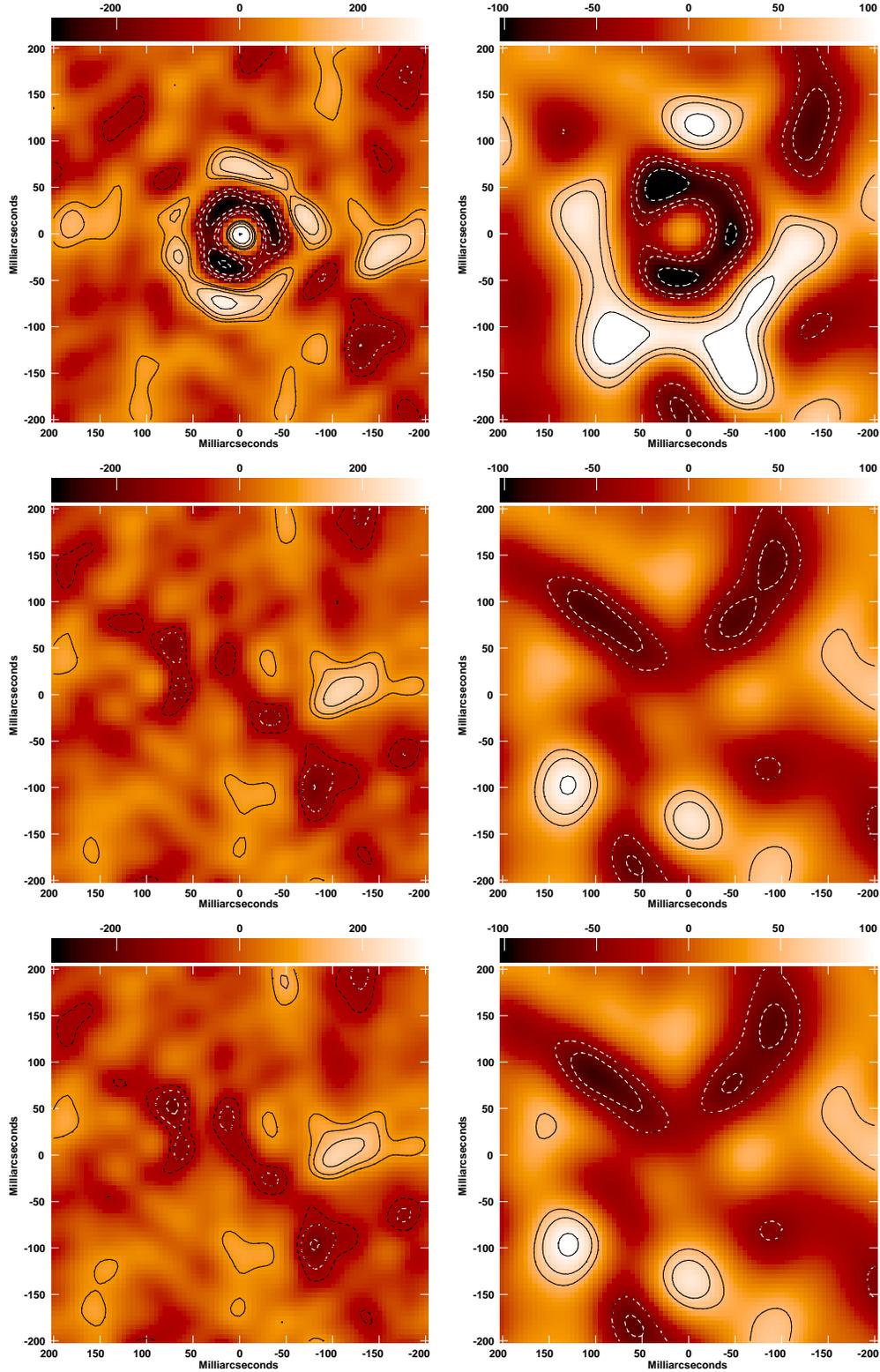

\centering
  \scalebox{0.35}{\rotatebox{0}{\includegraphics{f3a.ps}}}
  \scalebox{0.35}{\rotatebox{0}{\includegraphics{f3b.ps}}}
  \scalebox{0.35}{\rotatebox{0}{\includegraphics{f3c.ps}}}
  \scalebox{0.35}{\rotatebox{0}{\includegraphics{f3d.ps}}}
  \scalebox{0.35}{\rotatebox{0}{\includegraphics{f3e.ps}}}
  \scalebox{0.35}{\rotatebox{0}{\includegraphics{f3f.ps}}}
\caption{Residual maps obtained
  by imaging the $\lambda$7~mm data (left) and $\lambda$1.3~cm data (right) from 2019
  August 2, after
  subtraction of model fits from the visibilities. Top row: residuals from best-fitting
  uniform elliptical disk model (see Table~3); Middle row: residuals from a model comprising
  two uniform elliptical disk components (Model 1, Table~4); Bottom row: residuals from a model comprising 
  a uniform elliptal disk and elliptical ring (Model 2, Table~4). 
  Intensity units are $\mu$Jy beam$^{-1}$. Contours
  are [$-12,-8.5,-6,-4.2,-3, 3,...,12]\times\sigma_{\nu}$
where $\sigma_{\nu}$ is the 1$\sigma$ RMS noise in the respective band (Table~2). Negative contours are shown as dashed lines.}
\label{fig:residuals}
\end{figure}

If we subtract the best-fitting uniform
elliptical disk model from the visibility data and Fourier transform the
residuals to make an image, an outer ``ring'' of positive residuals is visible in both
the 7~mm and 1.3~cm data, surrounding a 
negative trough  and a positive
central peak coincident with the star's center
(Figure~\ref{fig:residuals}, top row). The peak residuals are
$\lsim12\sigma$ in the 7~mm image and $\lsim6.5\sigma$ in the 1.3~cm
image. 

The fits to both the 7~mm and 1.3~cm data are improved by the introduction of a
  second  component to the model. Results of
  a two-component elliptical disk model (Model~1) and an elliptical disk+ring model (Model~2)
  are shown on the upper panels of Figure~\ref{fig:visplots} as dashed
lines and dotted lines, respectively.  The model parameters are summarized in
Table~4. Both of these models result in a  significant reduction of the residuals in the image domain
(Figure~\ref{fig:residuals}, middle and lower panels),
although for the 7~mm data both models still underpredict the visibility amplitude at large
$u$-$v$ distances ($<5$M$\lambda$; Figure~\ref{fig:visplots}), with the discrepancy being larger for the disk+ring model.
At $\lambda$1.3~cm
the fit quality for Models~1 and 2 is similar.
We stress that neither of these two-component models is unique, nor do they have an obvious
underlying physical underpinning. However, it is noteworthy that because the two components are concentric, the presence of a large 
convective cell (cf. Lim et al. 1998) or other transient surface feature
cannot readily account for presence of the second component. 

It is unclear whether this multi-component brightness profile is  unique to our 2019
observations, as previous spatially resolved observations of Betelgeuse at comparable wavelengths were
obtained prior to the VLA's sensitivity upgrade
(Perley et al. 2011), making the unambiguous
identification of similar features more difficult. However, hints of deviation from a simple uniform
elliptical disk model are seen in fits to earlier data (cf. Figure~2 of O'Gorman et al. 2015).
In addition, O'Gorman et
al. (2017) also found a multi-component model necessary to fit their
2015 ALMA 338~GHz observations of Betelgeuse, and a  similar behavior was seen in 7~mm observations of
the nearby carbon
star IRC+10216 (Matthews et al. 2018). In general the
identification of such trends is important for constraining models
 of the electron temperature as a function of radius in red supergiant atmospheres
 (see, e.g., Figure~3 of Harper et al. 2001) and underscores the unique power of high-sensitivity, spatially
 resolved radio observations.

 \subsubsection{No Evidence for Giant Convective Cells\protect\label{cells}}
 Based on their 7~mm VLA observations, Lim et al. (1998)  reported a significant
 brightness asymmetry that they interpreted as possible evidence of a large
 convective cell in the atmosphere of Betelgeuse. They noted that such features
  may be linked with upwellings of
cool photospheric gas and may play a role in the mass-loss process of
red supergiants. Indeed, it has been suggested that the recent Great Dimming may have been caused
  by a mass-loss event tied to the surfacing of a giant convective cell (Montarg\`es et al. 2021).
  However, our VLA measurements from 2019 August do not reveal any direct evidence for the presence
of such features; we see neither morphological evidence of giant ``cells'' nor signatures of enhanced
radio emission.
  As discussed in Sections~\ref{params} and
 \ref{multicomp}, the star appears largely symmetric and uniform,
although the imaginary parts of the visibility data
(Figure~\ref{fig:visplots}, right panels) do show some
low-level fluctuations that may reflect small brightness
variations across the star. 
 One possible explanation is that any observable
effects of a photospheric convective cell within the regions
sampled by our observations had already
  disappeared after ``surfacing'' between
  2019 January--April.

\subsection{Discussion\protect\label{discussion}}
\begin{figure}
  \centering
  \scalebox{0.6}{\rotatebox{0}{\includegraphics{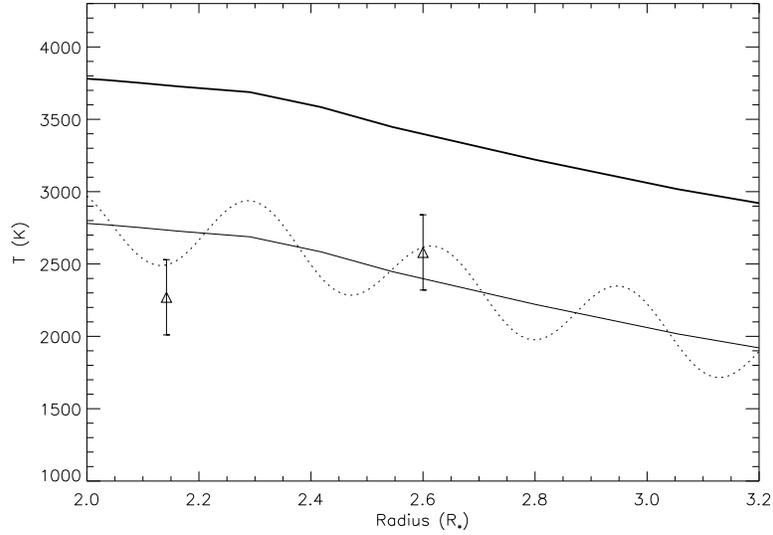}}}
  \caption{Schematic illustration of the effects of a propagating shock wave
    on the radial temperature profile of the extended atmosphere of Betelgeuse. The assumed
    shape of the underlying 
    temperature profile is based on
    the semi-empirical model of Harper et al. (2001). The thick solid line shows the temperature
    predicted by the original model; the thin solid line shows the same model with a temperature offset
    of $-$1000~K. The dotted line
    shows the temperature perturbations to the latter model resulting from a  periodic
    disturbance with $P$=400~days, amplitude $\Delta T_{s}=\pm250$~K, and
    propagation speed
   $V_{s}=$7.0~\kms\ (see text for details).   The triangle symbols
    indicate the brightness temperatures derived from the 2019 August 2 VLA measurements.
  }  
\label{fig:toymodel}
\end{figure}

As shown in
Figure~\ref{fig:toymodel}, the brightness temperature values that we derive from our
2019 August 2 observations of Betelgeuse at $\lambda$7~mm and $\lambda$1.3~cm lie significantly
below the predictions of the semi-empirical model of Harper et al. (2001).
The compendium of Betelgeuse radio measurements presented by O'Gorman et
al. (2015) shows that within the period from 1996--2004, the 1.3~cm flux density appears to have
systematically decreased by $\sim$20\% compared with the earlier measurements of Newell \&
Hjellming (1982) (which spanned 1972--1981) and Drake et al. (1992) (which spanned 1986--1990). The
physical mechanism that might explain such systematic dimming is unclear, although it may be linked with
long-term changes in the underlying dynamics of the star (e.g., George et al. 2020). 
To our knowledge, no radio light curve data for Betelgeuse are available between 2005 and mid-2019. Therefore
we are unable to determine conclusively whether
our present measurements  point to a continued,
systematic dimming trend of Betelgeuse at cm wavelengths, reflect normal temporal variations---or
alternatively, may be linked with atmospheric changes leading up to the UV outburst detected by
Dupree et al. (2020) and/or the subsequent Great Dimming. Here we briefly discuss this latter possibility.

As discussed in the previous section, our VLA observations of Betelgeuse on 2019 August 2 
 do not appear to show  any unambiguous
{\it morphological} signatures of an ongoing atmospheric disturbance in early
2019 August, roughly one month prior to the onset of the outflow
event documented by Dupree et al. (2020).
Despite this, our radio measurements are consistent with the possibility that the atmosphere of
Betelgeuse may have undergone marked physical changes near the time of
the outburst. Indeed, the anomalously low brightness
temperatures that we measure between $r\sim 2$--$3R{\star}$, as well as
the apparent temperature inversion within this radial
range, point to significant changes density
and/or ionization fraction within this region compared with previous epochs.

As briefly
described in Section~1, the tomographic analysis of Kravchenko et al. (2021) revealed evidence for the the
occurrence of successive shock waves through the photosphere of Betelgeuse in 2018 February and
2019 January, with the initial shock serving to amplify the second one.
Subsequently, spatially resolved spectroscopic 
measurements in the UV by Dupree et al. (2020) revealed that
a shock or pressure  wave passed through the southwestern portion of
Betelgeuse's chromosphere
between 2019 September and November.  (The exact time of the onset of this latter event is
not precisely known, but evidence for a disturbance was not seen during the
previous epoch of UV observations in 2019 March).  Concurrent measurements of the \CII\ lines at 2325 and 2328\ang\
also point to increases in the temperature and electron
density in the southern hemisphere of Betelgeuse. 
Importantly, the
UV measurements  of Betelgeuse's
chromosphere from Dupree et al. (including the \MgII\ lines and continuum)
probe approximately the same range of projected radii as our current VLA
observations ($\sim$2--3$R_{\star}$). These events are therefore also expected to impact
the radio continuum emission near the same time period.

As discussed by Reid \& Menten (1997), the propagation of pulsationally-induced
shocks through the extended atmosphere of a red giant ($r\gsim 2R_{\star}$) may produce temperature and density
variations as a function of radius that may in turn lead to variations in the
observed radio emission. 
In a simple version of this model, the temperature as a function of radius, $T(r_{0})$, predicted by a
hydrostatic model of the extended atmosphere becomes instead:

\begin{equation}
T(r_{0}) = \left<T(r_{0})\right> + \Delta T_{s}{\rm cos}\left(\phi_p + \phi_{s}\right)
\end{equation}

\noindent where $\Delta T_{s}$ is the amplitude of the temperature  contrast between the
pre- and post-shock gas, $\phi_{p}$ is the stellar phase,
and $\phi_{s}$ is the phase of the propagating disturbance. Following Reid \& Menten,
if one assumes
a shock propagation speed $V_{s}$ and a pulsation period $P$, then $\phi_{s}$ can be defined as

\begin{equation}
\phi_{s} = 2\pi\frac{r - R_{\star}}{V_{s}P}
\end{equation}

Figure~\ref{fig:toymodel} (dotted line) shows an example of the effect of this 
toy model on the radial temperature profile. We based the shape of the input temperature profile on
the semi-empirical model atmosphere of Harper et al. (2001), but have applied
a temperature offset of $-1000~K$ (thin solid line).
Here we have adopted $P$=400~days (comparable to Betelgeuse's ``short'' pulsation period; see Section~\ref{Intro}),
$V_{s}$=7.0~\kms, and $\Delta T_{s}$=250~K. We show the results for a single value of the shock excitation
phase, $\phi_{p}$=0.09. The exact combination of parameters is somewhat arbitrary. Nonetheless,
this toy model qualitatively illustrates that a disturbance of this kind may be able to account
for: (i)  a (temporary) distortion of the temperature profile, including a possible
temperature reversal between the radial regions sampled by the
7~mm and 1.3~cm radio emission, respectively;  (ii)  corresponding modulations  in the
radio flux densities; and (iii) temperature and electron density fluctuations across
the chromospheric regions probed by the {\it HST} UV
measurements of Dupree et al.
2020).
One important test of this scenario will be
follow-up observations
of Betelgeuse at radio wavelengths to assess whether the star has since returned
to its historical flux density levels.

%
\begin{deluxetable*}{lccllll}
\tabletypesize{\footnotesize}
\tablewidth{0pc}
\tablenum{4}
\tablecaption{Multi-Component Model Fit Parameters}
\tablehead{
\colhead{Comp.} &  \colhead{E offset (mas)} & \colhead{N offset
(mas)} & 
\colhead{$\theta_{a}$ (mas)} & \colhead{$\theta_{b}$ (mas)} & \colhead{PA (deg)} &
\colhead{$S_{\nu}$ (mJy)} }
\startdata

\tableline
\multicolumn{7}{c}{Model 1 ($\lambda$=7 mm)}\\
\tableline
Uniform Elliptical Disk & 0.25$\pm$0.22 & 0.09$\pm$0.21 & 125.6$\pm$1.6 &
116.5$\pm$1.3 &  $-33.6\pm1.7$ & 12.5$\pm$0.5 \\

Uniform Elliptical Disk & $-$0.24$\pm$0.16 & 0.05$\pm$0.13 & 78.3$\pm$0.1 &
69.1$\pm$0.1 &  $-93.7\pm1.2$ & 10.6$\pm0.6$ \\

\tableline
\multicolumn{7}{c}{Model 2 ($\lambda$=7 mm)}\\
\tableline
Uniform Elliptical Disk & $-$0.07$\pm$0.08 & 0.01$\pm$0.07 & 81.2$\pm$0.8 &
76.1$\pm$0.8 &  $95.6\pm1.1$ & 17.2$\pm0.6$ \\

Elliptical Ring & 0.29$\pm$0.26 & $-$0.01$\pm$0.03 & 110.5$\pm$1.4 &
101.6$\pm$1.1 &  147.6$\pm1.7$ & 5.9$\pm$0.2 \\

\tableline
\tableline
\multicolumn{7}{c}{Model 1 ($\lambda$=1.3 cm)}\\
\tableline
Uniform Elliptical Disk & 0.26$\pm$0.12 & 0.70$\pm$0.15 & 98.8$\pm$1.9 & 91.6$\pm$2.4 & 174.7$\pm$2.0 & 6.5$\pm$0.3\\

Uniform Elliptical Disk & $-1.20\pm0.49$ & $-3.6\pm0.63$ & 183.3$\pm$5.3 & 166.8$\pm$4.3 & 120.0$\pm$2.4 & 3.3$\pm$0.3\\
\tableline
\multicolumn{7}{c}{Model 2 ($\lambda$=1.3 cm)}\\
\tableline

Uniform Elliptical Disk & 0.03$\pm$0.08 & 0.06$\pm$0.09 & 98.9$\pm$2.0 & 92.9$\pm$2.2 & 77.5$\pm$1.7 & 7.7$\pm$0.2\\

Elliptical Ring & $-1.19\pm0.48$ & $-3.7\pm0.65$ & 151.1$\pm$4.1 & 137.3$\pm$3.3 & 26.9$\pm$2.2 & 2.1$\pm$0.2\\
\enddata
\tablecomments{Quoted error bars include only formal fitting uncertainties. The models are shown overplotted on the visibility
  data in Figure~\ref{fig:visplots}.}
\end{deluxetable*}

\section{Summary}
We have presented spatially resolved VLA imaging observations of the radio emission from Betelgeuse at
wavelengths of 7~mm and 1.3~cm. The data were obtained on 2019 August 2, just prior to  the onset of the Great
Dimming of the star in late 2019 and early 2020.
At both observed wavelengths the star appears 
spherically symmetric with no evidence of large-scale brightness non-uniformities. However, the
brightness profile in both bands is found to be more complex than a simple uniform elliptical disk.

We find the 7~mm flux density of Betelgeuse to be $\sim$20\% lower than previously published measurements during
the past 25 years.
The mean brightness temperature derived from the 7~mm observations, which probes gas at  a mean projected radius of
$r\sim2.1R_{\star}$,  is also significantly lower compared with previous measurements at comparable
radii
and lies $\gsim$1200~K below the prediction 
of published semi-empirical models of Betelgeuse's extended atmosphere. The brightness temperature derived
from the 1.3~cm data (which probes material at $r\sim2.6R_{\star}$) is also 
lower than expected based on trends in historical measurements. Our new data therefore suggest
recent changes in the density and/or temperature structure of the atmosphere between
$\sim 2$--$3R_{\star}$. One possible explanation is the recent passage of a large-amplitude shock
or pressure wave through the outer atmosphere.  This picture can also account for atmospheric disturbances revealed
by UV
imaging spectroscopy by Dupree et al. (2020) during the two months following the VLA observations. Such an
event may be linked to a large-scale mass ejection from the star that has been postulated as
an explanation for the steep decline in optical magnitude associated with the Great Dimming.

\bigskip

\acknowledgements
The observations presented here were part of NRAO program SQ0474. This work was supported by Space
Telescope Science Institute Grant HST-GO-15641.014-A. We thank G. Harper for helpful discussions.

%

%

%

\end{document}